\documentclass[conference]{IEEEtran}

\usepackage{booktabs}
\usepackage{paralist}
\usepackage[group-separator={,}, per-mode=symbol, binary-units]{siunitx}
\usepackage{color}
\usepackage{xspace}
\usepackage{eurosym}

\usepackage{graphicx}
\usepackage{color, colortbl}
\usepackage[ruled,vlined]{algorithm2e}
\usepackage{amsfonts}
\usepackage{threeparttable, tablefootnote}

\usepackage{multirow}
\usepackage{booktabs}
\usepackage{cite}
\usepackage{textcomp}
\usepackage[T1]{fontenc}%
\usepackage[export]{adjustbox}
\usepackage{placeins}

\newcommand{\ie}{i.\@\,e.\@\xspace}
\newcommand{\eg}{e.\@\,g.\@\xspace}

\newcommand{\etal}{et~al.\@\xspace}
\newcommand{\inch}{\,\textquotedbl}

\newcommand{\Paragraph}[1]{\smallskip\noindent{\bf #1.}}

\definecolor{Gray}{gray}{0.5}
\PassOptionsToPackage{hyphens}{url}\usepackage{hyperref}

\ifCLASSINFOpdf
\else
\fi

\usepackage{amsmath}

\ifCLASSOPTIONcompsoc
  \usepackage[caption=false,font=normalsize,labelfont=sf,textfont=sf]{subfig}
\else
  \usepackage[caption=false,font=footnotesize]{subfig}
\fi

\begin{document}

\title{Anti-Tamper Radio: System-Level Tamper Detection for Computing Systems}%
 
\author{\IEEEauthorblockN{
Paul Staat\IEEEauthorrefmark{1},
Johannes Tobisch\IEEEauthorrefmark{1}, 
Christian Zenger\IEEEauthorrefmark{2} and
Christof Paar\IEEEauthorrefmark{1}}
\IEEEauthorblockA{\IEEEauthorrefmark{1}Max Planck Institute for Security and Privacy, Bochum, Germany}
\IEEEauthorblockA{\IEEEauthorrefmark{2}PHYSEC GmbH, Bochum, Germany}
\IEEEauthorblockA{E-mail: \{paul.staat, johannes.tobisch, christof.paar\}@mpi-sp.org, christian.zenger@physec.de}
}

\maketitle

\begin{abstract}
A whole range of attacks becomes possible when adversaries gain physical access to computing systems that process or contain sensitive data. Examples include side-channel analysis, bus probing, device cloning, or implanting hardware Trojans. Defending against these kinds of attacks is considered a challenging endeavor, requiring anti-tamper solutions to monitor the physical environment of the system. Current solutions range from simple switches, which detect if a case is opened, to meshes of conducting material that provide more fine-grained detection of integrity violations. However, these solutions suffer from an intricate trade-off between physical security on the one side and reliability, cost, and difficulty to manufacture on the other.

In this work, we demonstrate that radio wave propagation in an enclosed system of complex geometry is sensitive against adversarial physical manipulation. We present an anti-tamper radio~(ATR) solution as a method for tamper detection, which combines high detection sensitivity and reliability with ease-of-use. ATR constantly monitors the wireless signal propagation behavior within the boundaries of a metal case. Tamper attempts such as insertion of foreign objects, will alter the observed radio signal response, subsequently raising an alarm.

The ATR principle is applicable in many computing systems that require physical security such as servers, ATMs, and smart meters. As a case study, we use $19$\inch~servers and thoroughly investigate capabilities and limits of the ATR. Using a custom-built automated probing station, we simulate probing attacks by inserting needles with high precision into protected environments. Our experimental results show that our ATR implementation can detect \SI{16}{\mm} insertions of needles of diameter as low as~\SI{0.1}{\mm} under ideal conditions. In the more realistic environment of a running $19$\inch~server, we demonstrate reliable detection of \SI{40}{\mm} insertions of needles of diameter~\SI{1}{\mm} for a period of $10$~days.

\end{abstract}

\section{Introduction}

\subsection{Motivation}
Today, almost all parts of our digital society require security mechanisms. These usually rely on cryptographic primitives and protocols which have been subjected to intensive scrutiny in the past decades. As a result, mathematical attacks are now largely infeasible and attackers often focus on other, weaker, links to break a system.
It is especially concerning when adversaries gain physical access to devices which opens a broad attack surface. Depending on the attacker's motivation, their goal might be to extract secret information, plant malicious functionality, or theft of intellectual property~(IP). %

There are a number of prominent examples for invasive attacks. For instance, the extraction of trusted platform module (TPM)-stored Bitlocker keys via bus sniffing has been shown~\cite{tpm_sniff_pulse,tpm_sniff_fsecure,tpm_sniff_dolos}. In this case, physical access completely unlocks the device\footnote{More secure Bitlocker configurations are available~\cite{ms_bitlocker}.}. As another example, the manipulation of PIN entry devices for electronic payments allows attackers to copy card data and PIN codes~\cite{drimerThinkingBoxSystemLevel2008}. In another case, it was possible to circumvent the encryption of FPGA bitstreams~\cite{enderUnpatchable2020}, allowing IP theft and injection of malicious circuit alterations. A real world example with far reaching consequences can be found in the Snowden documents. The NSA~\cite{spiegel_NSA} is apparently able to intercept and backdoor hardware in transit in a manner that makes it virtually impossible for the recipient to detect tampering. This could, for example, result in trojanized routers and firewalls which give persistent access to otherwise secure networks.

To defend against these kinds of attacks, a system designer can add tamper-detection solutions that monitor the physical state of the system. Any detected attack raises an alarm and the system usually has to resort to ``self-destruction'', which includes the wiping of any sensitive information such as secret key material before the attacker has a chance at extraction. While anti-tamper techniques do not provide the same level of security as expected by standards of modern cryptography, they are still an important building block for the protection of real-world devices. In practice, employed solutions range from  simple switches, that detect if a case has been opened, to complex hardware security modules~(HSM). These use sophisticated meshes made of conducting material as distributed sensors of physical integrity and environmental sensors, e.g., for temperature or light, and many additional measures.

The most important certification standards for devices that require physical security are FIPS 140-2~\cite{nistFIPS140_2}, its recent successor FIPS 140-3, and Common Criteria~\cite{commonCriteria}. 
 In the case of FIPS, four levels of security are defined. Starting from the third level, compliant devices are expected to detect physical attacks and to react to them by zeroization of critical security parameters (CSPs). A database of devices certified in the NIST's cryptographic module validation program is publicly available~\cite{nist_cmvp_database}. As an example, the Amazon AWS Key Management Service HSM provides level three physical security. Its description defines ``the cryptographic boundary [...] as the secure chassis of the appliance'' and says that ``all key materials are maintained exclusively in volatile memory in the appliance and are erased immediately upon detection of physical tampering''~\cite{nist_cmvp_aws_hsm}. Further documentation~\cite{nist_cmvp_aws_sec_policy} only reveals that ``the module’s production-grade enclosure is made of a hard metal, and the enclosure contains a removable cover'' and that ``[...] an internal tamper switch zeroizes CSPs at power on / power off when triggered [...]''. In contrast, the IBM~4767 cryptographic coprocessor, that comes in form of a PCI express card, is certified at level four. In its documentation, it is said that ``physical security is constantly monitored through a tamper detection/ response envelope with tamper response and zeroization circuitry''~\cite{nist_cmvp_ibm_sec_policy}. This is achieved by a security mesh, voltage and temperature sensors.  A predecessor of this system, the IBM 4758, was analysed by Anderson~\cite{andersonSecEngineering} and was found to provide good security against physical attacks.

The downsides of existing solutions are that they either lack in security against sophisticated attackers, as in the case of rather simple tamper switches, or that they are costly and inflexible. The second part is highlighted by the elaborate manufacturing process of IBM HSMs, as detailed in~\cite{isaacs2013tamper}, which consists of multiple intricate steps and intensive quality control. Furthermore, using the existing techniques, it is difficult to extend tamper detection to the system level, e.g., to a complete server case. Especially in this case, one has to contend with reliability issues due to aging and environmental changes.

\subsection{System Idea}
\label{sec:system_idea}

\begin{figure}
\centering
\includegraphics[width=0.6\linewidth]{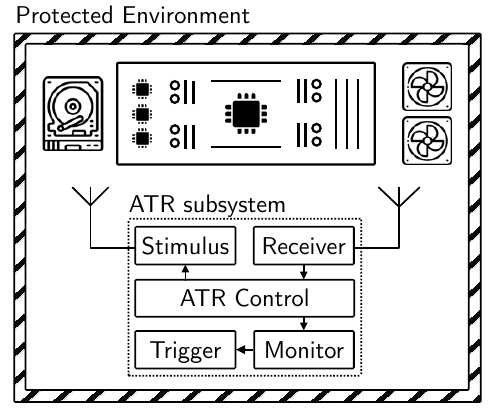}
\caption{Anti-tamper system overview. Stimulus and receiver form the physical layer and measure a response, i.e., a ``fingerprint'' of the surrounding radio channel. When the monitor registers a large deviation to the reference, an alarm is triggered.}
\label{fig:system_overview}
\end{figure}

In this work, we examine the use of radio waves as a tool for tamper detection. Phenomena of wave propagation, such as reflection and scattering, make radio waves sensitive to their physical propagation environment, which can be utilized in security applications as addressed in previous work. %
A prominent example is the use of radio measurements to sample entropy from the environment by two communicating parties in order to derive shared cryptographic keys~\cite{zhang_key_2016}. For this application, changes in the environment are considered a useful stochastic process. In contrast, other applications focus more on the fact that radio measurements can be regarded as fingerprints of their environment, i.e., static parts of the channel response are the focus. One example can be found in indoor localization schemes~\cite{chenWifiIndoorLocalization}. Here, the radio channel identifies the position of a device in relation to base stations. This idea of fingerprinting can be straightforwardly transferred to the security context. For example, DeJean and Kirovski~\cite{dejeanRFDNARadioFrequencyCertificates2007} presented hardware tags whose authenticity could be verified by a near-field measurement. Bagci~\etal~\cite{bagciUsingChannelState2015} described a scheme for the detection of tampering of IoT devices that are connected to trusted nodes via WiFi. Tampering was experimentally carried out by manually moving and rotating devices which could be detected by analysis of channel state information from multiple receivers. In a recent talk, Zenger~\etal~\cite{enc_puf_36c3} proposed monitoring the state of an enclosed system with radio waves. A tamper event, i.e., insertion of a foreign object into the device, should alter the radio wave propagation enough to be detected.

In our work, we follow this general idea. We assume that the target system that is to be protected is enclosed in a metal case. A set of antennas is embedded to measure the radio wave propagation behavior within the case, i.e., the wireless channel response. Changes in the protected environment such as opening the case, insertion of probing needles, movement of cables, introduction of additional ICs (``mod chips''), leave an imprint on the measured radio channel and thus triggers an alarm. The relation between observed channel and the environment is complex, i.e., computing an exact mapping in either direction is practically infeasible. This prevents an attacker from adjusting the environment to compensate for the effects of their intrusion.

In the following, we refer to a system protected by this physical mechanism as an anti-tamper radio (ATR) solution.

There are several advantages to this approach. For one, it is very flexible. No potentially brittle security mesh needs to be wrapped around the device. Instead larger systems with complex geometry can be covered with just a few antennas. This even allows for the possibility of retrofitting existing devices and is less costly to implement than traditional solutions. In contrast to security meshes, which cannot be opened and sealed again, a wireless approach offers better accessibility and allows re-initialization of the legitimate state.

\subsection{Contributions}
In this work, we provide an experimental evaluation of a prototypical ATR implementation. In particular, we put an emphasis on the reproducible simulation of attacks, which is not found in the literature so far. Our contributions are as follows:

\begin{compactitem}
    \item We present a custom-built automated probing station that can simulate attacks by inserting needles with high precision into protected environments.
    \item We examine an experimental ATR system in two such environments. The first one, an empty metal case, presents an idealized system with minimal noise. We use this to investigate basic tamper detection capabilities and show that our system detects even needles of diameter \SI{0.1}{mm} reliably, when inserted at a depth of at least \SI{16}{\mm}. 
    \item As our second target, we use a regular $19$\inch~server that presents a more realistic scenario with noise sources such as fans and temperature swings. In this harsher environment, we demonstrate the impact of legitimate changes such as variation of temperature and computing load of the server. We achieve long-term stable tamper detection over $10$~days.
    \item For most of our radio measurements we use a versatile but costly vector network analyzer (VNA). Additionally, we also show that ATR measurements can be realised with cheap, commercial off-the-shelf hardware.

\end{compactitem}

\section{Technical Background}

In this section, we provide technical background information on tamper-proof systems and wireless sensing.

\subsection{Tamper-Proof Hardware}
The protection of devices, \eg, sensitive computing systems, against unauthorized physical access is referred to as \textit{tamperproofing}. The scope of protection can range from individual parts to modules and entire devices, \eg, from a single chip to circuit boards and entire systems. Especially the protection of larger and less stringently sealed systems is considered an open issue. Anti-tamper solutions typically fulfill one or multiple of the following key functionalities.

While not directly providing hurdles to attackers, \textit{tamper evidence} prevents tamper events from going undetected upon inspection of physical integrity by an honest party. Tamper evidence is usually achieved by means of materials with special visual characteristics that are irreversibly deformed upon tampering. Examples include seals, potting, or coatings~\cite{appelSecuritySealsVoting2011, andersonSecEngineering}. \textit{Tamper resistance} aims at hampering physical access, \ie, by providing strong system boundaries that make tamper attempts cumbersome and thereby less attractive as the attacker faces increased efforts in time or cost. Examples include bank safes and potting of printed circuit boards~(PCBs). Devices that implement a \textit{tamper detection} mechanism may recognize a tamper event as it occurs. Tamper detection necessarily involves sensor readings to provide an interface to the physical world. Here, sensors detect anomalous system behavior such as, \eg, the breach of an conductive envelope, incident light, vibrations, or temperature deviations~\cite{weingartPhysicalSecurityDevices2000a}. Tamper detection mechanisms are implemented to trigger a \textit{tamper response}. Depending on the value of the protected device and the CSPs, the response could result in complete physical self destruction or wiping of sensitive data or key material.

Manufacturers of physical security solutions to date still tend to pursue security-by-obscurity principles for their products~\cite{helfmeierBreakingEnteringSilicon2013, andersonSecEngineering}, which is why detailed system internals or attacks are barely available to the public.

\subsection{Wireless Sensing}
As wireless radio signals travel from a transmitter to a receiver, they are affected by their propagation environment. That is, effects such as multipath signal propagation make the received signals carry information about their surroundings. \textit{Wireless sensing} seeks to extract such information by analyzing received signals. Dedicated wireless sensing systems are well-known and established. For instance, radar systems can detect objects and individuals for applications such as speed control, aircraft monitoring, or human gesture recognition~\cite{lienSoliUbiquitousGesture2016}. More recently, wireless sensing is also implemented with standard wireless communication systems~\cite{paulSurveyRFCommunications2017}, \eg, \mbox{Wi-Fi}~\cite{maWiFiSensingChannel2019} and ultra wideband~(UWB)~\cite{qionghuangUWBThroughWallImaging2010}. That is particularly attractive as such solutions can provide multi-objective functionalities beyond mere communication. Furthermore, the channel estimates required for wireless sensing are produced in any case as they are an essential part of reliable wireless communication. This is done in the receiver's low-level baseband signal processing by analyzing the (known) preamble portion of received signals. %
Applications such as, \eg, human activity and gesture recognition, imaging, or vital sign monitoring~\cite{yousefiSurveyBehaviorRecognition2017, maWiFiSensingChannel2019}, impressively demonstrate the ability of commodity wireless devices to detect environmental conditions. Notably, ongoing standardization efforts for an upcoming \mbox{Wi-Fi} standard IEEE~802.11bf, specifically geared towards wireless sensing, are expected to be finalized in fall~2024~\cite{ieee_80211bf}.

\section{System Outline and Adversary Model}

\begin{figure}
\centering
\includegraphics[width=1.0\linewidth]{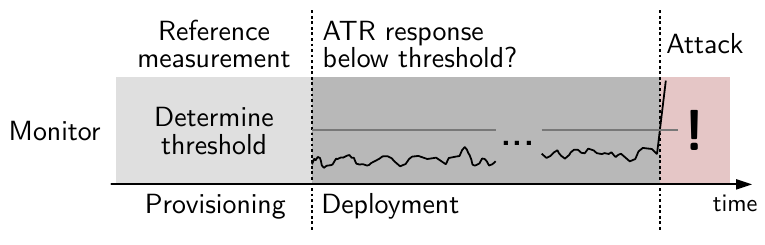}
\caption{Illustration of ATR phases}
\label{fig:system_phases}
\end{figure}

As briefly outlined in Section~\ref{sec:system_idea}, we make the assumption that the ATR protects a system that is enclosed in a metal case that reflects radio waves well (see \autoref{fig:system_overview} for an overview). The case houses a computing base~(CB) that contains critical security parameters, \eg, secret data or intellectual property. The ATR itself is designed as a subsystem to the CB and is part of the protected environment. It consists of antennas, a measurement system and a security monitor that keeps track of deviations in the measured signals. Upon breach detection, \ie, a physical tamper event, the security monitor informs the CB, which acts by deleting confidential data. The integrity is continuously live-checked and requires a battery-backed power supply. A (temporary) loss of power must be treated as a loss of integrity because the monitoring system itself could have been tampered with and can no longer be trusted.

The antenna placement, measurement hardware and selected frequency bands all govern which type of intrusion can be detected. Especially for complex geometries, it is possible that some regions are not well covered by radio waves. We term these non-sensitive regions, in contrast to sensitive regions which typically are found close to and in line-of-sight of the antennas. It is upon the system designer to decide if non-sensitive regions can be tolerated or need to be covered by additional antennas.

In theory, tamper events generated by an attacker can take many forms. In this work, we concentrate on a specific type of attack, namely the insertion of a conductive needle into the system. This insertion models an adversarial attempt to probe electric signals that carry sensitive information. Our attacker model, thus, defines the defeat of the ATR as the ability of the attacker to perform a needle insertion into a sensitive region without being detected.

Focusing on this specific form of tampering allows us to conduct reproducible experiments in different configurations. In particular, we can scale the extent of the attack by varying the needle size and insertion depth. Furthermore, such rather subtle tampering subsumes other classes of manipulations. If a needle of diameter \SI{0.3}{\mm} can be detected, then surely opening lids or insertion of larger objects will be detected as well.

Besides the sensitivity against even small physical manipulations, another major requirement for the application of an ATR is long-term stability. In its simplest form, the ATR collects a reference measurement during provisioning. Once deployed, the ATR then continuously performs measurements and compares these against the reference, as illustrated in~\autoref{fig:system_phases}. However, due to different sources of noise in the measurement process, the measurements will always exhibit a certain difference to the reference that needs to be tolerated. Additionally, the system needs to withstand legitimate environmental changes without sounding the alarm, such as temperature variations or the movement of fans within the system, all of which have an impact on the radio wave propagation within the enclosure. Therefore, it is important to characterize the system's stability over time including the impact of legitimate changes and thereby setting a threshold for the detection of insertion attacks.

\section{Experimental Setup}

We now describe our automated probing attack testbed, the test enclosures and radio measurement hardware.

\subsection{Experimental Probing Setup}
\begin{figure}
\centering
\includegraphics[width=0.9\linewidth]{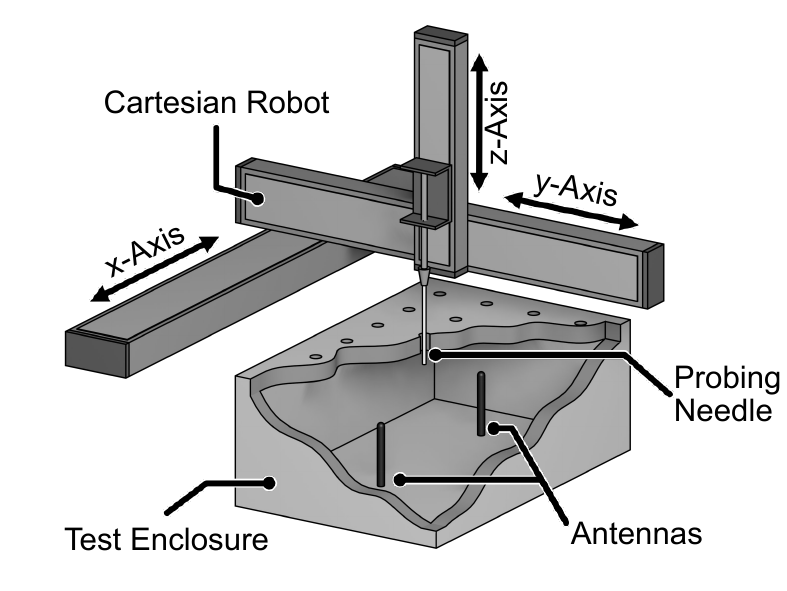}
\caption{The basic experiment setup consists of a test enclosure that acts as a stand-in for a protected environment. The antennas placed within the enclosure are connected to external radio equipment not shown here. A Cartesian robot lowers a probing needle through holes in the enclosure lid. }
\label{fig:exp_render}
\end{figure}

\begin{figure}
\centering
\includegraphics[width=1.0\linewidth]{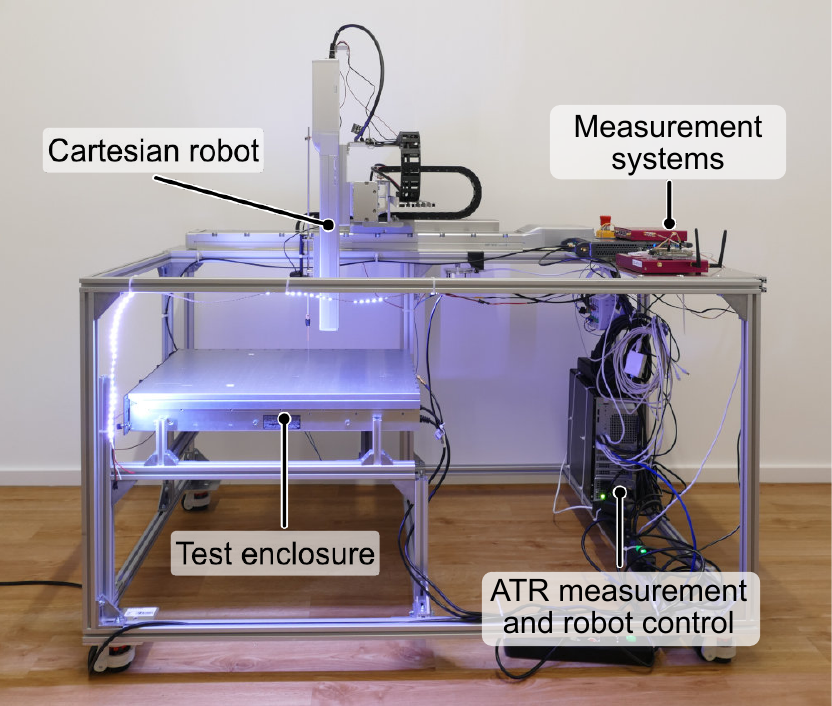}
\caption{Experimental setup consisting of a Cartesian robot for probing needle positioning, a test enclosure, radio measurement systems, and experiment control.}
\label{fig:experimental_setup}
\end{figure}

\begin{figure}
\centering
\includegraphics[width=1.0\linewidth]{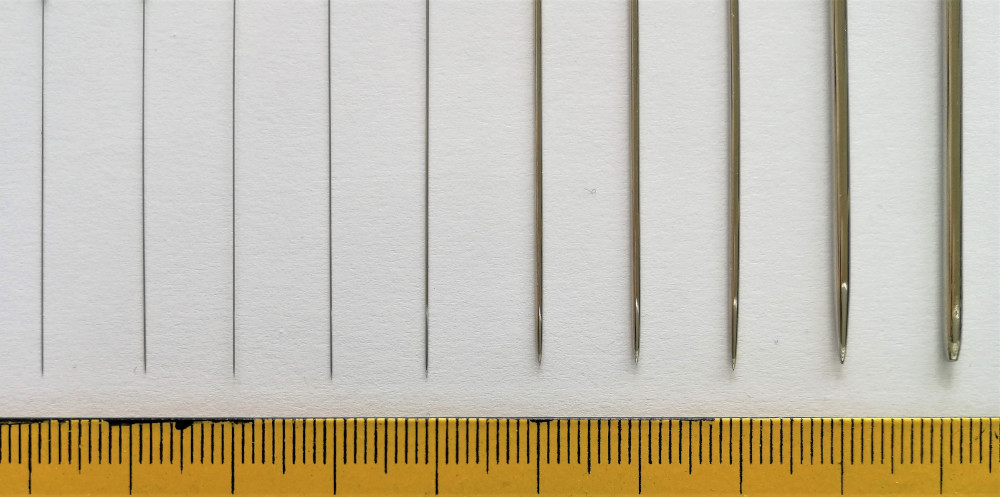}
\caption{Close view of several probing needles of diameter \SI{0.16}{\mm}~(left) to \SI{2}{\mm}~(right). Scale at the bottom shows millimeter.}
\label{fig:needles}
\end{figure}

At the core of our experimental setup is an automated needle positioning system that can produce tamper events with high repeatability~(see \autoref{fig:exp_render} and \autoref{fig:experimental_setup}). It consists of an aluminum strut frame on which a commercially available three-axis Cartesian robot and a test enclosure are mounted. Each of the robot axes is independently controllable with a resolution of \SI{0.01}{\mm}. A probing needle is mounted on a slide on the z-axis. By moving in the x- and y-directions, the needle is positioned relative to the enclosure and by lowering the z-axis it is inserted into the test enclosure. 

For the probing, we use metallic needles for medical applications, which are available at reasonable cost and in different sizes. The needles have lengths from \SI{3}{\cm} to \SI{10}{cm} and diameters between \SI{0.1}{\mm} and \SI{2}{\mm}. A selection of needles is depicted in~\autoref{fig:needles}. %

\subsection{Test Enclosures}
\label{sec:test_enclosures}
We investigate the ATR system within two test environments on which we carry out simulated physical tampering attacks. The first is an empty metallic box with which we mimic an idealized computer housing. The second test environment is a fully functional $19$\inch~server.

\begin{figure}
\centering
\includegraphics[width=1.0\linewidth]{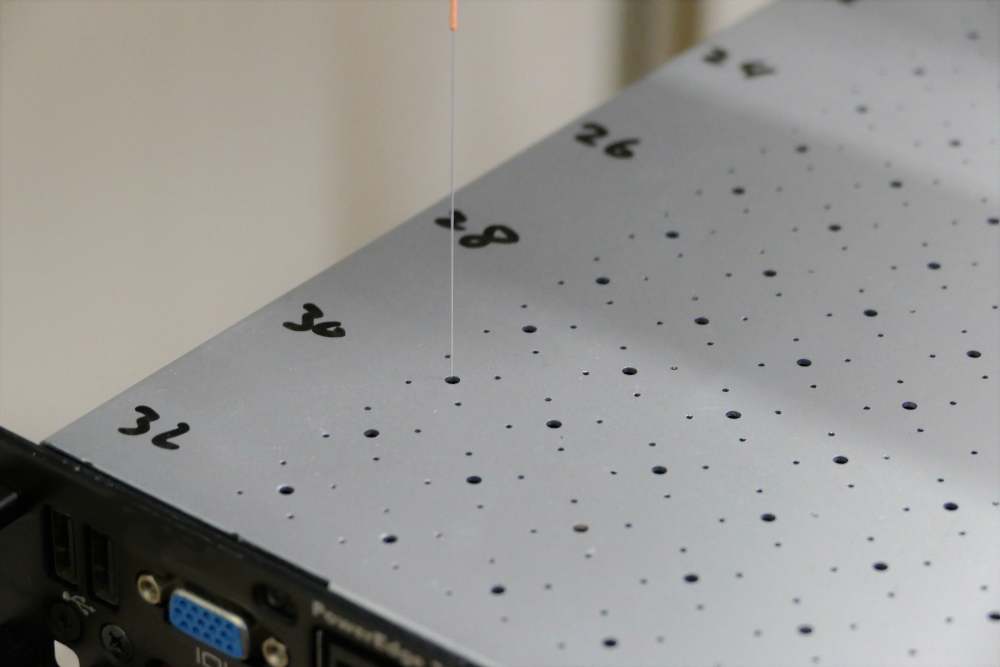}
\caption{Close view of the server lid with drills and probing needle.}
\label{fig:server_lid}
\end{figure}

We drilled holes in a grid pattern into the top lids of both enclosures, as shown in~\autoref{fig:server_lid}, which allows us to simulate reproducible tamper events, \ie, we insert metallic needles through these holes. Note that the holes of diameter \SI{2.5}{\mm} are rather small compared to the radio wavelengths in the order of \SI{3}~to~\SI{14}{\cm} that we leverage for the ATR measurements. Therefore, the holes have negligible effect on the RF shielding properties of the cases (we also demonstrate this experimentally). %

\Paragraph{Idealized Box}
The use of an empty aluminium box allows us to exclude potential noise sources that would be present in an actual computer case, \eg, from fan movement, temperature variation, or electromagnetic interference. The dimensions of the box are \SI{60}~$\times$~\SI{31}~$\times$~\SI{11}{\cm}. We have installed two SMA coaxial feedthroughs on the enclosure walls so that we can connect antennas to our measurement systems from the outside. The lid of the metallic box contains $416$~holes designated for our probing attacks. The drills are arranged in a rectangular grid with a distance of \SI{2}{\cm}.

\Paragraph{Server Unit}
As a case study for a real-world application of the ATR, we use a fully functional Dell~Poweredge~2850 server running Ubuntu. The unit is a $19$\inch~device for rack mounting with a height of $2$~rack~units (dimensions are \SI{45}~$\times$~\SI{76}~$\times$~\SI{9}{\cm}). The server lid has $636$~holes, however due to the internal structure of the device, only $170$~holes are suited for needle insertion. The server has two quad-core Intel Xeon CPUs running at \SI{3}{\GHz} and has \SI{4}{GB} of RAM. We use the Linux program \emph{stress} to apply computing load variations. In particular, we can vary the number of utilized cores, producing fast temperature swings within the server case. Moreover, we use \emph{ipmi-sensors} to obtain sensor readings indicating temperatures and fan speeds of the server.

\subsection{Measurement Systems}
We have tested two RF measurement systems for our experimental investigation of the ATR. To make experimentation easier, the measurement hardware is placed outside the test enclosures and connected via coaxial cables to two wideband antennas~\cite{uwbant_manual} within the enclosures. An electronically controlled RF switch~\cite{rfswitch_manual} allows us to select between the different measurement systems. Each of these systems yields a real or complex-valued channel vector to which we refer to in the following as the system's \emph{response}. For our prototypical ATR implementation, we leverage a block averaging over $10$~consecutive responses.

In the following, we outline the used measurement systems. With the described configurations, we obtained satisfactory results regarding noise level and measurement speed.

\Paragraph{Vector Network Analyzer}
As the main measurement system we use a Keysight P9372A vector network analyzer~(VNA) that is capable of complex channel measurements in the frequency domain up to \SI{9}{\GHz}~\cite{p9372a_manual}. The main application of a VNA is the scattering parameter characterization of two-port RF circuits such as amplifiers or filters. Therefore, a VNA combines a transmitter and a receiver in a single device. It applies a stimulus signal (an unmodulated RF carrier) to the device under test~(DUT) and measures the signals being reflected and transmitted by the DUT. In order to obtain the frequency response, the VNA repeats this procedure rapidly while varying the carrier frequency. In this manner, the VNA is likewise capable of acquiring the radio channel between two antennas as it is required for ATR. 

In our experiments, we connect the two antennas within the test environment to the two ports of the VNA to obtain the magnitude transfer function\footnote{We measure the forward transmission magnitude from VNA port~$1$ to port~$2$, $|S_{21}(f)|$, using an IF bandwidth of \SI{2}{\kHz}.}, covering $500$ equally spaced frequency points between \SI{2}{\GHz} and \SI{9}{GHz}. For the ATR, we apply smoothing in frequency direction using a simple moving average filter. Obtaining a single response takes approximately \SI{250}{\ms}.

Unless otherwise noted, we performed experiments with this device.

\Paragraph{Ultra-Wideband Radio}
Besides the VNA, which we consider as the ground truth for our study, we additionally tested cheap off-the-shelf measurement systems based on conventional UWB radio transceivers. We obtained satisfactory results using commercially available and fully integrated UWB transceivers~\cite{dw1000_manual}. These are available at less than US\$5 and implement the IEEE 802.15.4 UWB physical layer based on the transmission of short radio impulses. We use two UWB transceivers, that are connected to the antennas within the test enclosure, to transmit and receive short packets\footnote{Channel bandwidth \SI{500}{\MHz}, data rate \SI{6.8}{Mbit/s}, 512 symbols preamble length, pulse-repetition frequency \SI{64}{\MHz}.}. %
As a by-product of the wireless communication, the receiver provides a fine-grained channel impulse response~(CIR) estimation at approximately \SI{1}{\ns} resolution. The CIR describes the radio channel from a time domain perspective, \ie, individual multipath components can be distinguished over arrival time at the receiver. We extract $15$ taps of the magnitude CIR after the first path component, corresponding to a maximum multipath length of approximately \SI{4.5}{\m}. Using a modified control software to run the transceiver outside of its core specification, we are able to gather CIR measurements on $11$~consecutive frequency channels between \SI{2.496}{GHz} and \SI{7.488}{GHz}, covering a total bandwidth of \SI{5.5}{\GHz}. Obtaining a single response over the entire bandwidth takes approximately \SI{700}{\ms}.

\subsection{Comparing Channel Measurements}
In an ATR system, we face the challenge of identifying anomalous measurements that occur upon violation of the environment's physical integrity, \eg, when an attacker inserts a foreign object into the enclosure. Here, we introduce the metric that we used to evaluate our measurements.

We assume that the ATR response $H_k[t]$ at time~$t$ is a real-valued vector\footnote{The original responses of our measurement systems are complex-valued but we have discarded all phase information because these are typically subject to severe measurement imperfection.} of $L$ elements. Individual elements are indexed with the subscript $k$. $H_k[t]$ may represent a channel transfer function, \ie, a VNA measurement with $k$ denoting frequency, or a CIR, \ie, an UWB measurement with $k$ denoting time-delay. Now, to assess the similarity between a pair of measurements at a single index $k$ but at different points in time, say the current measurement $H_{k}[t]$ and a reference measurement $H_{k}[t_0]$, we leverage the following metric from the literature~\cite{doerryMeasuringChannelBalance2018}:
\begin{equation}
    d_k(t, t_0) = 1 - 2\,\frac{\sqrt{|H_k[t]|^2 \ |H_k[t_0]|^2}}{\left( |H_k[t]|^2 + |H_k[t_0]|^2 \right)}.\label{eq:channel_distance}
\end{equation}
This normalized distance becomes 0 when both responses are equal and approaches 1 if both responses are very different. The normalization is required because elements of the response vectors along the index $k$ can take values that are multiple orders of magnitude apart. A regular absolute distance or squared distance would thus disregard parts of the spectrum that have a low amplitude but still show significant differences between the two responses. Thanks to the normalization, we can aggregate the distance between two response vectors to a single scalar by taking the mean over the per-element distances $d_k$. We refer to this measure as the \textit{mean normalized deviation}~(MND):
\begin{equation}
    \textrm{MND}(t, t_0) = \frac{1}{L} \sum^{L-1}_{k=0} d_k(t, t_0). \label{eq:average_distance}
\end{equation}

\section{Results} 
Any tamper-detection system must contend with the fact that there is a fundamental trade-off between sensitivity and accuracy of tamper detection. If one sets the detection threshold too low, the system is marred by false-positives which lead to unnecessary system shutdowns or even completely bricked devices. On the other hand, if one increases the margin of deviation from the reference too much, attacks will potentially be missed. In this section, we show our experimental results that shine light on this trade-off for ATR systems. 
We start by examining a baseline of what manipulations can be detected in ideal circumstances, i.e., what kind of needle insertions have a measurable effect in a simplified environment. Then, we gradually move towards a more realistic scenario, first by changing the reflectivity within the environment and then by adding noise sources in the forms of fans and temperature deviations. With appropriate signal processing, the system can detect our mock attacks even in this difficult scenario. Lastly, we also show that our professional measurement equipment can be replaced with cheap commercial-off-the-shelf hardware, albeit with some loss in detection accuracy.

\subsection{Probing Detection in Idealized Box}
\label{sec:probing_ideal_box}

\begin{figure}
\centering
\includegraphics[width=0.95\linewidth]{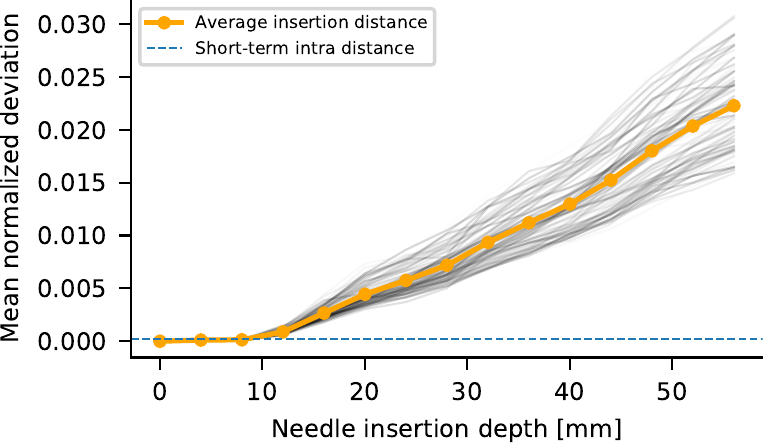}
\caption{Average Euclidean distance of ATR responses against the insertion depth of a \SI{0.3}{\mm} needle. For completeness, we also show results for individual positions in grey.}
\label{fig:idealized_depth}
\end{figure}

\begin{figure}
\centering
\includegraphics[width=0.95\linewidth]{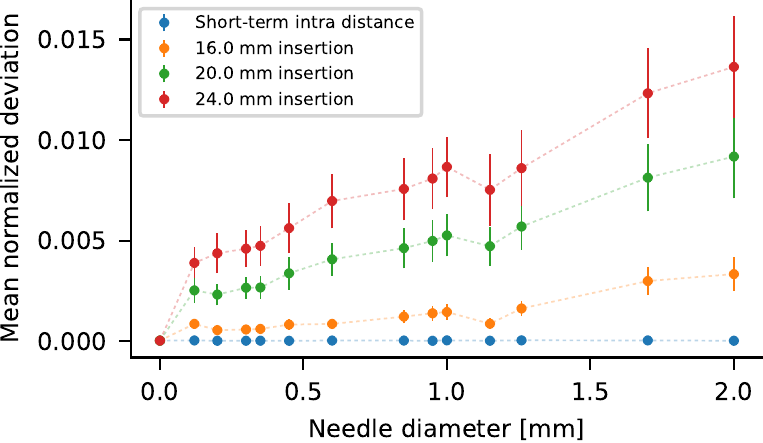}
\caption{Impact of the needle thickness on the detectability, indicating mean and standard deviation over the considered needle positions. A diameter of zero means that no needle was mounted (but the robot still moved).}
\label{fig:idealized_thickness}
\end{figure}

In our first experiment, we determined the impact of the needle insertion depth. To this end, we iteratively measured the system response for the insertion of a needle of \SI{0.3}{\mm}~diameter into a subset of $84$~holes (equidistantly sampled out of the total set of $416$~holes). In detail, the measurement process is as following. First, the needle is positioned in x~and~y~direction above the hole and a response is captured twice with the needle being completely outside the box. We call the difference between these two measurements the \textit{short-term intra distance}. This is the best-case measurement error governed by the noise found in the measurement system, as virtually no drift in environmental condition occurs in this short time-span. Afterward, the needle is gradually lowered in \SI{4}{\mm} steps down to a total depth of \SI{56}{\mm}. After each step, a response is captured. In \autoref{fig:idealized_depth}, we show the average insertion distance between the reference measurement, when the needle is outside, and the measurement with inserted needle. This is contrasted with the average short-term intra distance. It can be seen that a noticeable decorrelation only sets in after an insertion of around \SI{10}{\mm}\footnote{In the experiments, the needle was connected to a microcontroller pin. We have verified that the potential of the needle does not impact the detectability by the ATR, see Figure~\ref{fig:needle_potential} in the Appendix.}. Starting from there, the insertion distance rises approximately linear with the insertion depth. This experiment already confirms one basic ATR principle: minute physical changes can be detected by radio measurements. 

Before putting this claim to test in a more realistic scenario, we repeated the experiment with needles of varying diameter, from \SI{0.1}{\mm} up to \SI{2}{\mm}. In order to speed up the experiment, we reduced the number of measured holes to~$21$ that were spread out along the main axis between the antennas. The result is given in \autoref{fig:idealized_thickness}, where the insertion distance is shown for three depths. The curves confirm that the greater physical extent of the needles directly leads to better detectability, as the insertion distance grows approximately linearly with the needle diameter\footnote{Apart from needle insertions, Figure~\ref{fig:lidopening} in the Appendix shows that large physical changes such as opening a lid affect the ATR response even stronger.}. The figure also shows the effect of only moving the robot without a mounted needle (data points at diameter equalling zero). The distance is virtually the same as the short-term intra distance. This shows that the movement of the robot alone does not impact the radio response noticeably, as previously claimed in~\autoref{sec:test_enclosures}.

\subsection{From Idealized Box to Realistic Enclosure}
\label{sec:ideal_box_to_realistic}
As already noted, the empty aluminum box is only an idealized environment with different physical properties from a realistic enclosure. It can be understood as some sort of reverberation chamber -- a highly reflective environment wherein signals decay rather slowly. Thus, after excitation of a signal, many delayed copies will arrive at the receiver due to the multipath signal propagation within the enclosure. This circumstance aids the detection of physical manipulations which affect a large number of multipath components.

\begin{figure}[t]
\hspace*{\fill}%
\subfloat[]{%
        \includegraphics[width=0.3\columnwidth, angle=180]{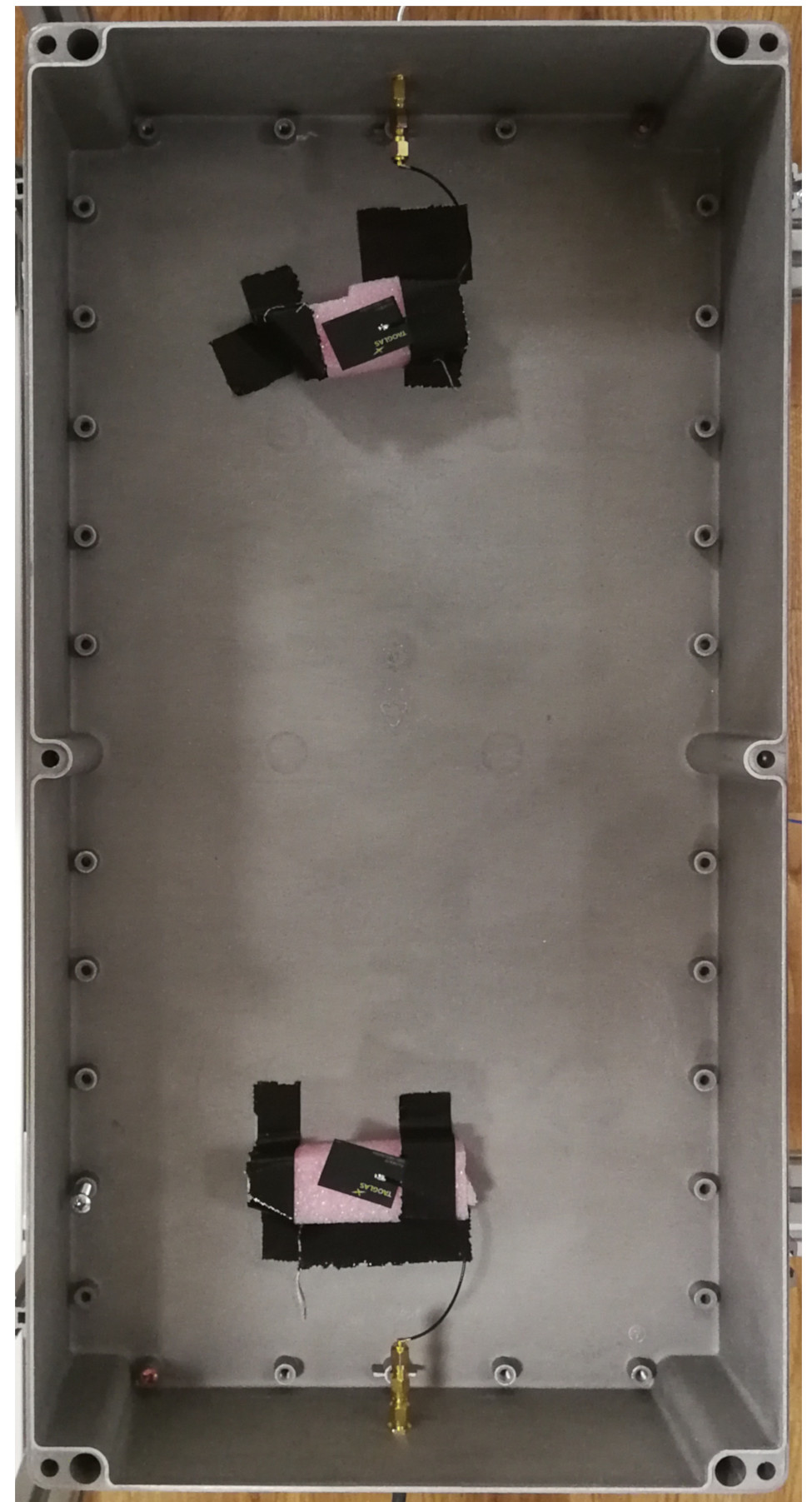}
        }
\hfill
\subfloat[]{%
        \includegraphics[width=0.3\columnwidth, angle=180]{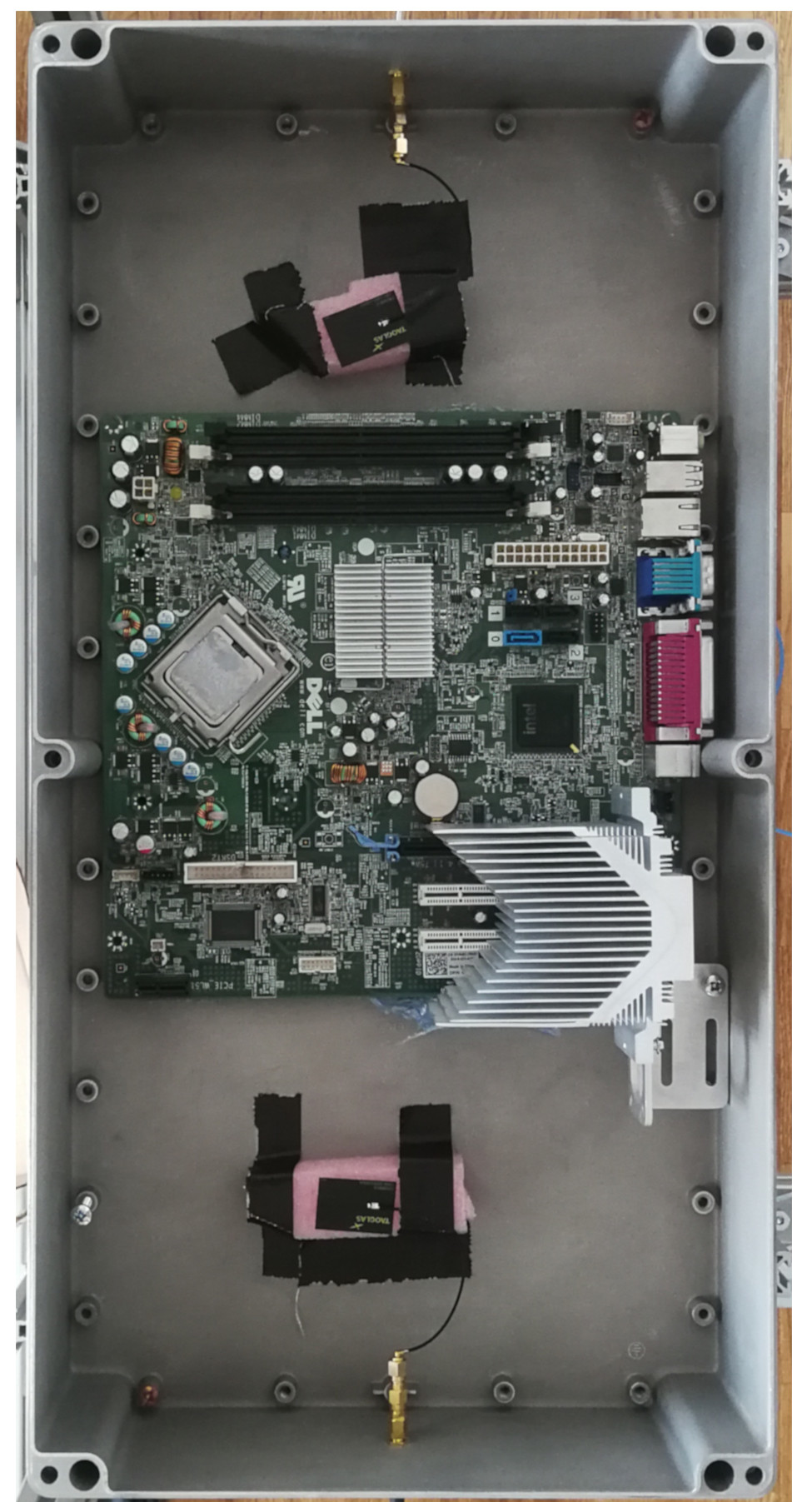}
        }
\hfill
\subfloat[]{%
        \includegraphics[width=0.3\columnwidth, angle=180]{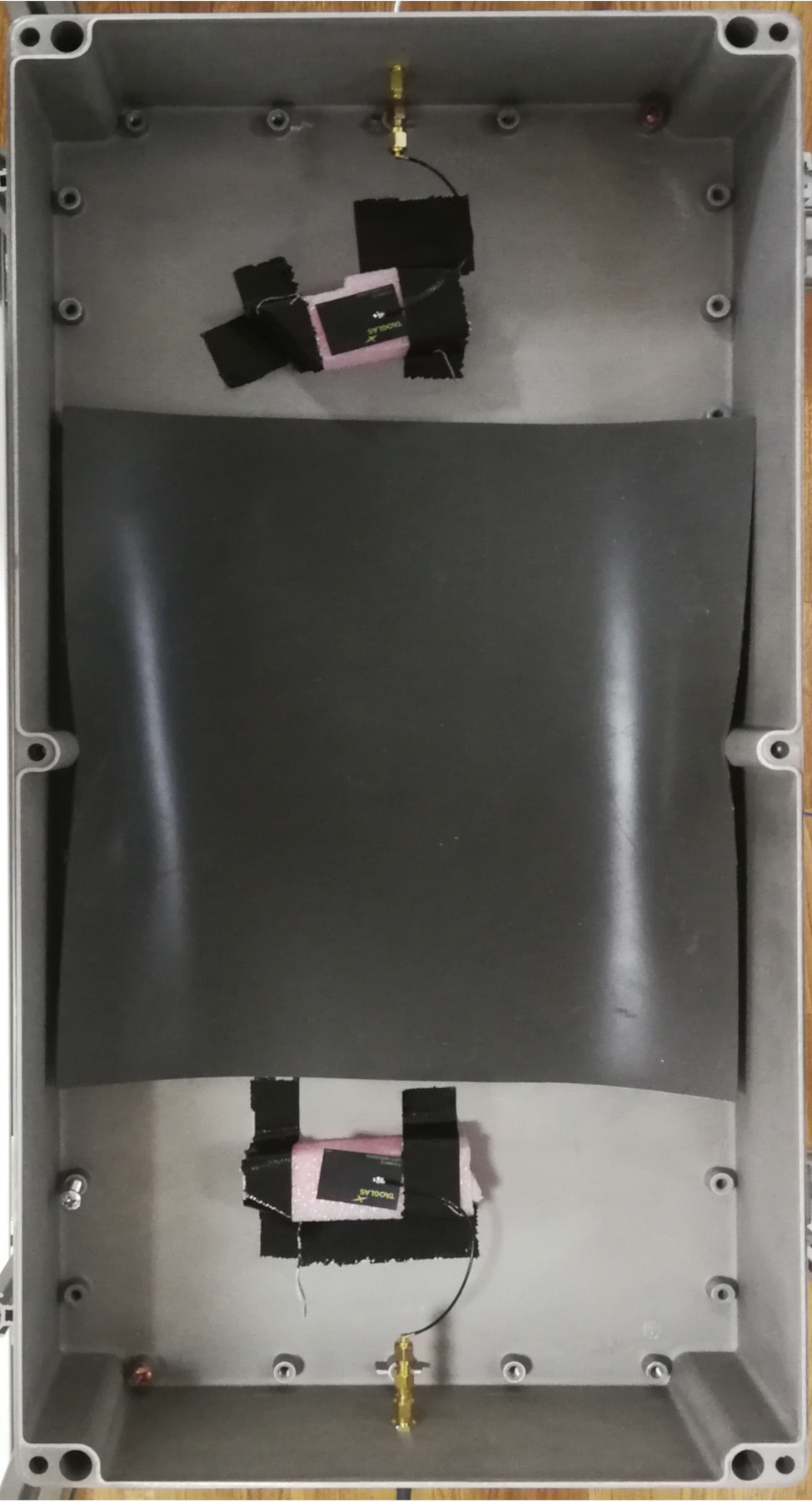}
        }
\hspace*{\fill}%
\caption{Inside view of the aluminium box with antennas and different interiors: (a)~empty, (b)~PCB and blocking metal object, (c)~RF absorber.}
\label{fig:interior_fotos}
\end{figure}

To investigate this effect, we ran a needle probing experiment with three different box configurations as shown in \autoref{fig:interior_fotos}: ($a$)~empty, ($b$)~a computer mainboard within the box, and ($c$)~an RF~absorber within the box. The mainboard is non-functional but acts as a passive object that alters radio wave propagation by introducing, \eg, absorption, shadowing, and additional reflections. The RF~absorber is a plastic mat of dimensions \SI{30}{\cm}~$\times$~\SI{30}{\cm} that is specifically designed to absorb radio waves in the GHz~frequency range~\cite{absorber_reference}.

\begin{figure}
\centering
\includegraphics[width=0.95\linewidth]{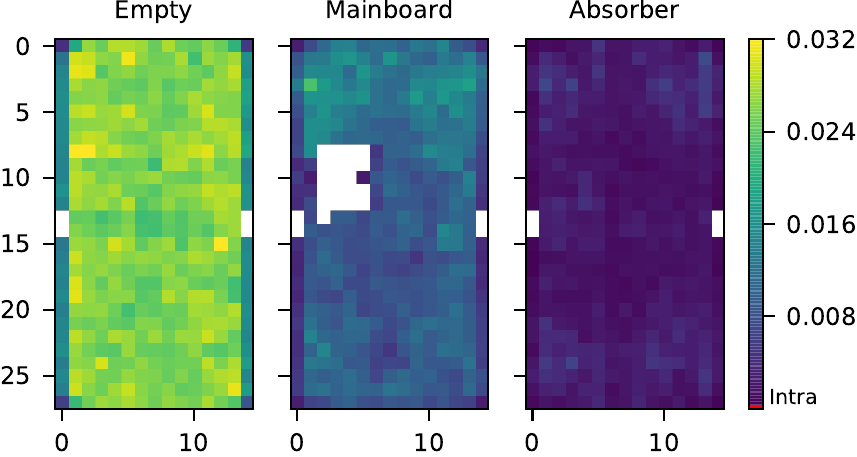}
\caption{Impact of a \SI{0.3}{\mm} diameter needle inserted \SI{45}{\mm} into the aluminium box for each position. The color scheme indicates the MND of responses with and without the needle inserted.  Darker blue color means less sensitivity against probing, while bright yellow spots show the highest sensitivity. %
Results for three box configurations: empty, loaded with a computer mainboard, loaded with an RF absorber.}
\label{fig:interior_heatmaps}
\end{figure}

\begin{figure}
\centering
\includegraphics[width=0.95\linewidth]{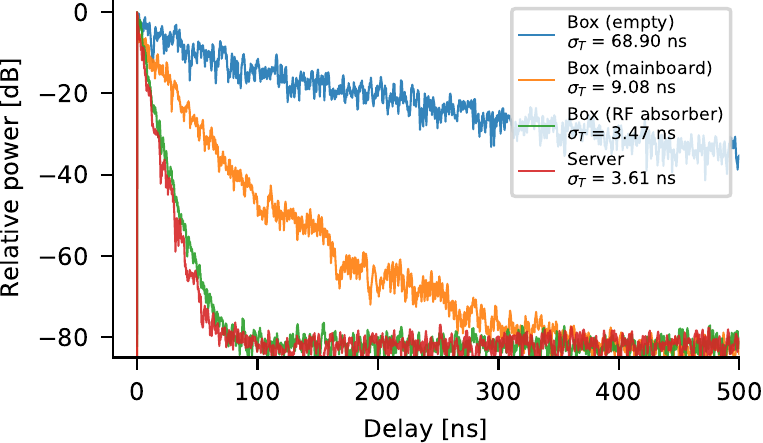}
\caption{Power delay profiles of the server and the metallic box with different loading. We also indicate the root mean square~(RMS) delay spread $\sigma_T$~\cite{goldsmith_wireless_2005}, an important metric to characterize channel time dispersion.}
\label{fig:enclosure_vs_server_pdp}
\end{figure}

In the experiment, we inserted a \SI{0.3}{\mm} diameter probing needle \SI{45}{\mm} into each drilled hole of the enclosure. In \autoref{fig:interior_heatmaps}, we show heatmaps for the three configurations, indicating the MND due to the needle insertion for each position. The key observation to make is that the needle detection margin is gradually reduced as the reflectivity found within the box is lowered by the mainboard and the absorber. %
A clearer picture of this is given by the power delay profiles in \autoref{fig:enclosure_vs_server_pdp}. Here, we again collected responses with the VNA for the three different box configurations and additionally a response in the more realistic server environment (in unpowered state), this time without any probing attempts. The power delay profile essentially shows the power of different multipath components over the arrival time at the receiver and is obtained through an inverse Fourier transform of the frequency response delivered by the VNA~\cite{goldsmith_wireless_2005}. Due to the nature of radio wave propagation, later signal components experience increasing attenuation, as they traverse longer through the enclosure. It can be clearly seen that the empty box, by far, has the most reverb, as expected. The mainboard already attenuates the signals to some degree, while absorber and the server enclosure show an even stronger (and surprisingly similar) behavior. This already indicates that we can expect a lessened detection performance in realistic scenarios, even without additional factors such as running fans.

\begin{figure}
\centering
\includegraphics[width=0.95\linewidth]{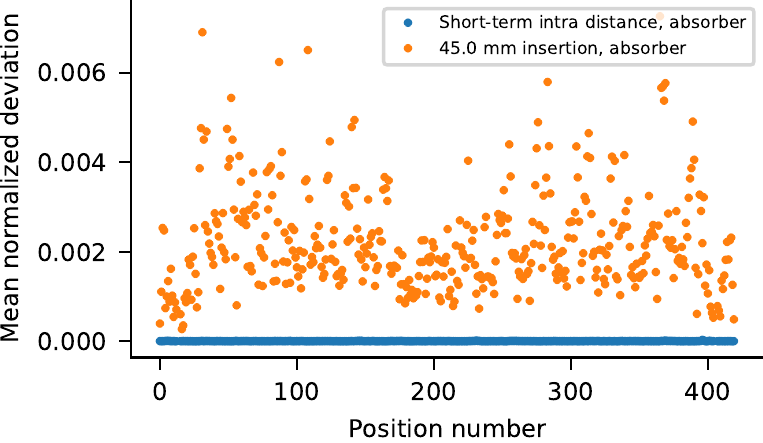}
\caption{Short-term intra distance and impact of \SI{45}{\mm} insertion of an \SI{0.3}{\mm} diameter needle for each hole of the aluminium box with RF absorber. We plot the flattened results, \ie, along the number of individual probing positions.}
\label{fig:absorber_only_all_positions}
\end{figure}

However, one should not be misled by the color scheme of the heatmaps and the seemingly bad performance when the box is loaded with the RF absorber. Despite the reduced detection margin, the needle is still detectable on each position. This is also shown in \autoref{fig:absorber_only_all_positions}, where we plot the impact of the needle insertion and the intra distance for each of the probing positions of the aluminium box with the absorber in place.

Another observation to be made in~\autoref{fig:interior_heatmaps}, is the factor that the border regions next to the walls show the needle insertion less clearly than the inner regions. This is due to the electromagnetic field distribution within the enclosure. Most importantly, a detection of the needle demands field components to be present. As discussed in the literature, regions within a distance of a quarter wavelength to the enclosure walls may be considered insensitive~\cite{serraReverberationChambersCarte21}. We made a similar observation in \autoref{fig:idealized_depth} where the inserted needle is only detected from a depth of \SI{10}{\mm} onward. The important lesson from that is that not all regions within an enclosure are necessarily equally well protected. For an actual system, it is thus an engineering problem to identify sensitive parts of the system and to ensure that these are well covered by the ATR. In our server test case, we probe from the top while the server mainboard is mounted close to the bottom side. Thus, the bottom side of the mainboard would presumably be in a region of low detectability and be open to attacks, which could only be solved by a system re-design.

\subsection{Noise Model}
In the previous section, we have already seen that the ability of detecting probing attempts can be impaired by the internal structure of the enclosure. In this section, we explore the impact of additional noise sources. For instance, in our server test case, the environment will behave rather dynamic: Several fans operate constantly, load variations lead to fast temperature variation, and a server certainly produces electromagnetic interference~(EMI).

\newcommand{\effChannel}{H^{(e)}_k[t]}
\newcommand{\trueChannel}{H_k[t]}
\newcommand{\envNoise}{N_k^{(env)}[t]}
\newcommand{\measNoise}{N_k^{(meas)}[t]}

We model the effective measured radio channel $\effChannel$ as the the sum of three parts
\begin{equation}
\effChannel = \trueChannel  + \measNoise + \envNoise.
\end{equation}

The first part, $\trueChannel$, is the true ideal channel response. The second component is the measurement noise, $\measNoise$. This is the thermal noise inherently present at the receiver and is the main contributing factor to the short-term intra distance that we have already encountered in \autoref{sec:probing_ideal_box}. The third factor, $\envNoise$, is the environmental noise, that encompasses all physical changes such as fan movements, vibrations, or thermal extension. We assume that $\measNoise$ is zero-mean while $\envNoise$ may drift over time.

\begin{figure}
\centering
\includegraphics[width=0.95\linewidth]{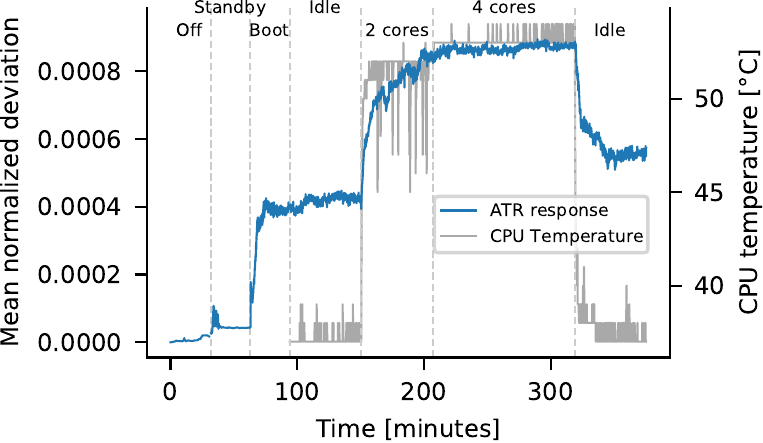}
\caption{Impact of server operation on measurements.}
\label{fig:intra_server_states}
\end{figure}

We give an example for this in \autoref{fig:intra_server_states}, for which we have captured responses from the server enclosure over time and over different states. First, we measured a reference response in a completely powered-off state. Then we plugged in power, which activates the power supply fans. Subsequently, we booted the server and added CPU load, in between two idle phases. It can be seen that the response stays relatively stable per state. The addition of noise source, such as power supply and then main cooling fans, leads to steep changes in response distance between states. The response does not simply drift away but can also come back closer to its initial state, as evident by the second idle phase.

\subsection{Probing Attacks on a Realistic Target}
\begin{figure}
\centering
\includegraphics[width=0.99\columnwidth]{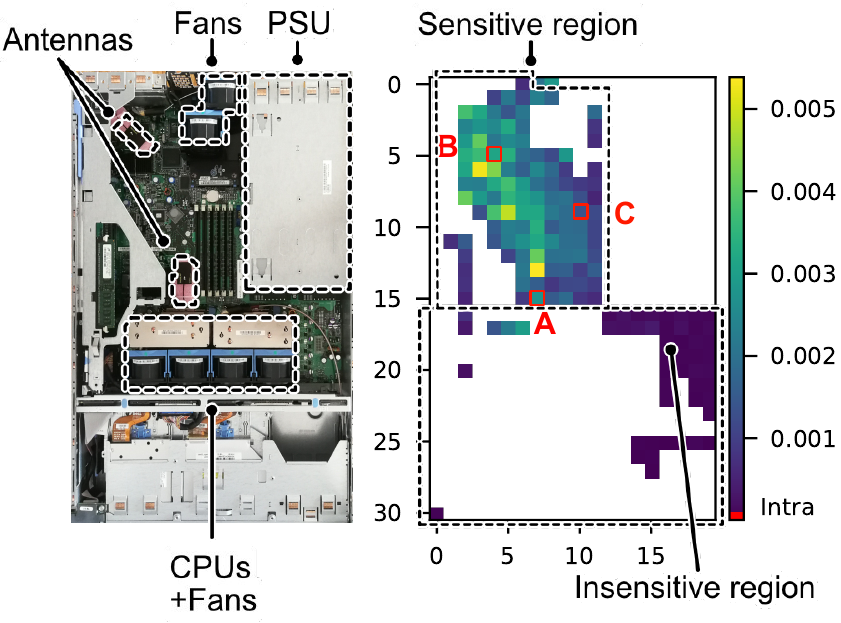}\caption{Server top view and location-dependent needle impact on the ATR response. The color scheme indicates the MND of responses with and without the needle inserted. Darker blue color means less sensitivity against probing, while bright yellow spots show the highest sensitivity. Note that the power supply unit (PSU) has additional fans that are not visible.}
\label{fig:server_areas}
\end{figure}

After our first study of server's the noise behavior, we started investigating how well needle intrusion could be detected. The first thing to note is that the geometry within the server case is rather complex (see left side of~\autoref{fig:server_areas}). Only some regions are accessible for vertical probing at all. Our first experiment was targeted at finding out how well these were covered by our ATR. To this end, we first disabled noise sources such as fans and lowered a needle of diameter \SI{1}{\mm} to a depth of \SI{40}{\mm} into the case. As shown in the right of Figure~\ref{fig:server_areas}, we could broadly identify two types of region, a (largely) sensitive one and an insensitive. The latter simply does not contribute strongly to the radio channel response as it is mostly shaded and therefore not well-covered by the antennas. Additional antennas are required if one wished to protect it. In all following experiments, we have limited our probing runs on the sensitive region consisting of a total of $117$~holes\footnote{Please note that we have excluded two holes at the top of the sensitive region to prevent uncontrolled needle movement due to fan airflow.}.

Next, we performed a long-term experiment featuring server load variations and repeated probing attacks (needle diameter: \SI{1}{\mm}, insertion depth: \SI{40}{\mm}) for $10$~days. During this time, we applied CPU load variations alternating between 0\% and 100\% with a period of approximately three hours. The deviation to the reference response can be found in Figure~\ref{fig:longterm_measurement_vna_raw}. The first thing to note is the strong periodicity of the long-term intra distance which directly reflects the changes in CPU load. Likewise, the insertion distance increases and decreases along the intra distance, i.e., physical changes caused by legitimate variations and by tamper events add up to a certain degree which is in line with our assumed noise model. In general, the needle insertions at most of the probing positions can be distinguished from normal states of operation. However, this yet sub-optimal detection performance can be substantially improved by selecting stable parts of the ATR response, as discussed in the next section.

\subsection{Spectrum Selection and Long-term Stability}
\begin{figure}
\centering
\includegraphics[width=0.95\linewidth]{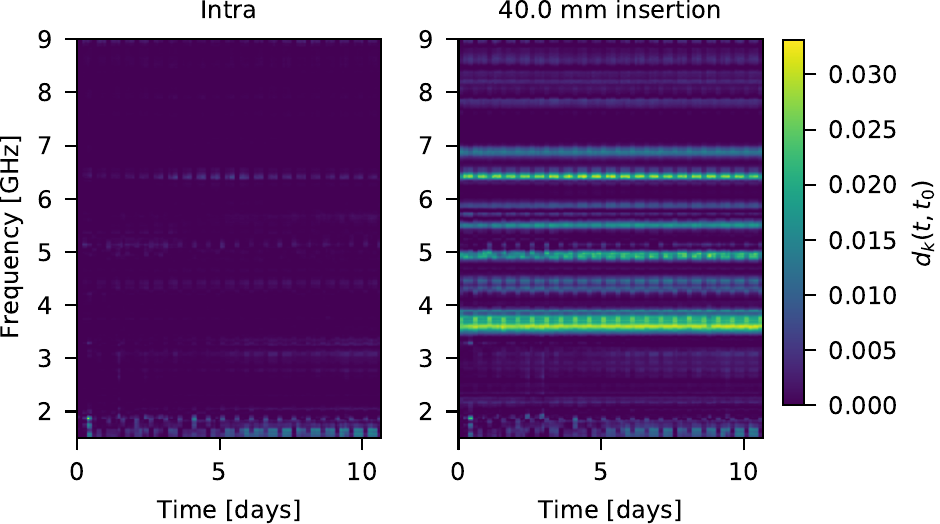}
\caption{Spectrogram of channel distances of ATR responses to the initial reference measurement for intra measurements and a \SI{40}{\mm} needle insertion at a single position.}%
\label{fig:spectral_effect_needle}
\end{figure}
So far, we have compared responses with the aggregated distance from~\autoref{eq:average_distance} that summarizes the complete spectrum into a single scalar. However, it is also possible to examine which parts of the spectrum contribute the most to either the intra distance or the insertion distance. In \autoref{fig:spectral_effect_needle}, we have done exactly this for the previously mentioned long-term experiment which also included server load variations. Here, we compute the channel distance between measured frequency responses over time to the reference measurement using~\autoref{eq:channel_distance}. %
Furthermore, we distinguish between untampered (intra) and tampered states and plot the results as spectrograms. The tampered state corresponds to a \SI{40}{\mm} insertion of the probing needle at a single position. It can be seen that the needle at this specific position leaves a spectral imprint that ($i$)~is well distinguishable from the intra measurements ($ii$)~while the decisive spectral regions in both states do not overlap significantly and are stable over time. Another observation is that ($iii$)~the periodic server load variation is only recognized in parts of the measured intra spectra. The first aspect is mandatory for a needle detection to be possible at all. From the other observations, however, we conclude that we can exclude parts of the spectrum that contribute to the intra distance without sacrificing the needle detection significantly. Such parts of the spectrum may exhibit a low signal-to-noise~(SNR) or an overly strong sensitivity to legitimate environmental changes. In the depicted spectrograms, such regions are found, for instance, below \SI{2}{\GHz} and around \SI{6.5}{\GHz}.

\begin{figure}
\centering
\includegraphics[width=0.95\linewidth]{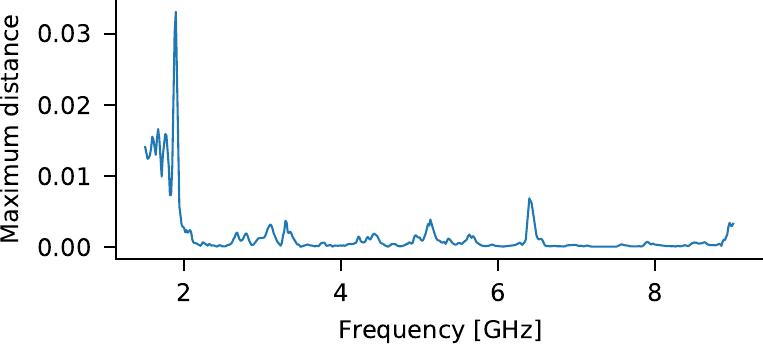}
\caption{Maximum channel distances $\alpha_k$ to the reference measurement over frequency, indicating spectral regions fluctuating due to noise and \emph{not} due to violation of physical integrity.}%
\label{fig:maximum_distance}
\end{figure}

Based on these findings, we have devised a preprocessing scheme that leverages a short provisioning phase of the ATR, which  allows us to identify fluctuating spectral regions from a set of $M$~legitimate ATR responses. For this, we use the same procedure as we did for obtaining the spectrograms in~\autoref{fig:spectral_effect_needle}, \ie, we use~\autoref{eq:channel_distance} to compute the distances of the initial reference response (captured at point in time $t_0$) to the responses gathered during provisioning (captured at points in time $t$, with $t>t_0$). Then, we gather the maximum distance over the $M$ responses, \ie, $\alpha_k = \textrm{max}\{d_k(t + m, t_0) : m = 1,\ldots,M\}$. Please note that this maximum distance is computed per frequency. For illustration purposes only, we evaluate the intra distances from~\autoref{fig:spectral_effect_needle} in this manner and plot all $\alpha_k$ in~\autoref{fig:maximum_distance}. Here, we can see that $\alpha_k$ is appropriate to identify the fluctuating regions below \SI{2}{\GHz} and around \SI{6.5}{\GHz} as the strongest contributors to the intra distance. 
After the provisioning phase, we select \SI{30}{\percent} of the largest values of $\alpha_k$ and drop the corresponding frequencies from the response evaluation during the deployment \ie, reducing the length of the ATR responses to $0.7 \, L$. We want to stress that the preprocessing takes no information about tamper events into account and only collects responses that are known to be legitimate. This is important because it means that the provisioning can be done without subjecting each device to individual tamper tests.

\begin{figure*}[!t]
\centering
\subfloat[]{\includegraphics[width=0.95\columnwidth]{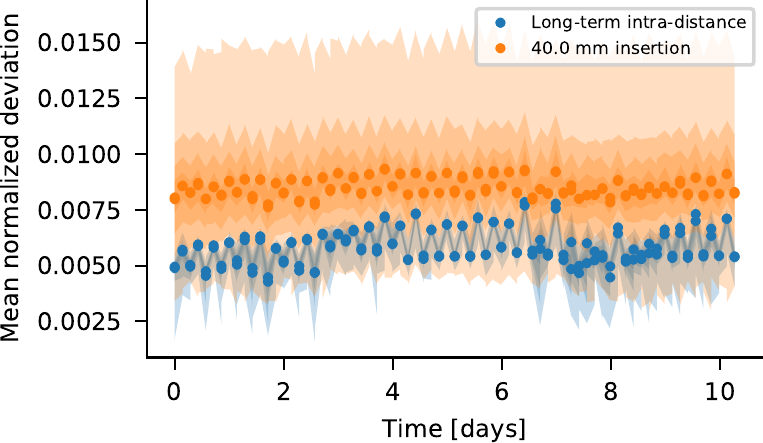}%
\label{fig:longterm_measurement_vna_raw}}
\hfill
\subfloat[]{\includegraphics[width=0.95\columnwidth]{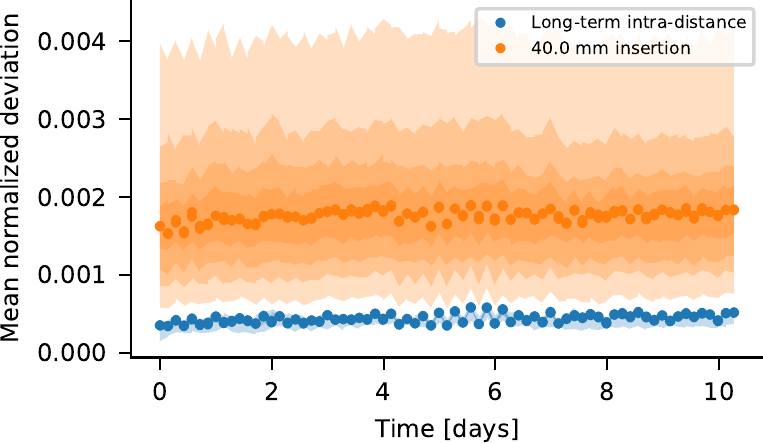}
\label{fig:longterm_measurement_vna}} \\
\subfloat[]{\includegraphics[width=0.95\columnwidth]{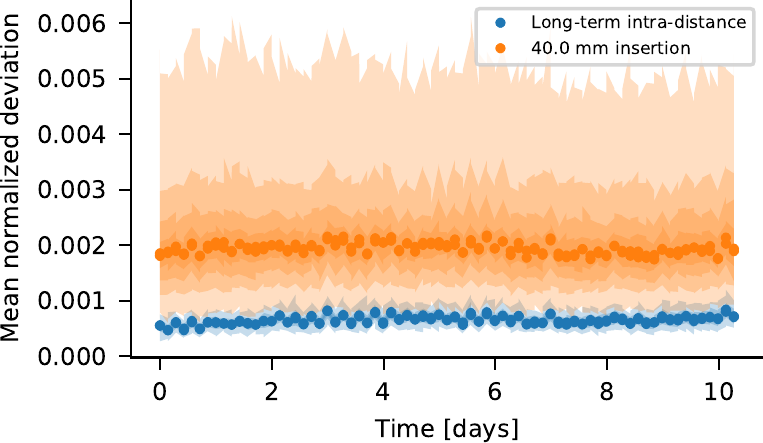}
\label{fig:longterm_measurement_uwb}}
\hfill
\subfloat[]{\includegraphics[width=0.95\columnwidth]{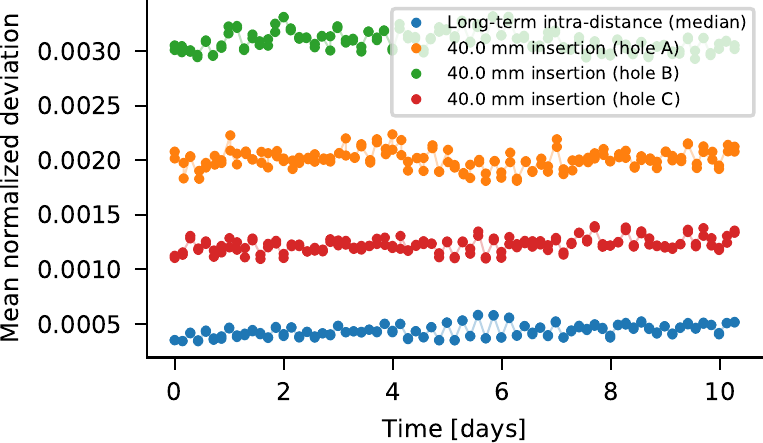}%
\label{fig:single_drill_plots}}
\caption{Longterm probing experiment with needle diameter of \SI{1}{mm} and different types of response. (a) VNA without preprocessing. (b) VNA with preprocessing. (c) UWB. (d) VNA with preprocessing. Selected holes only, see~\autoref{fig:server_areas} for the locations. - Please note that only distances in (b) and (d) stem from the same spectral data. The other figures can only be compared qualitatively. The shaded areas cover \SI{25}{\percent}, \SI{50}{\percent}, \SI{75}{\percent}, and \SI{99}{\percent} of the distribution of insertion distances over the $117$~tested holes.} 
\label{fig:longterm_measurement_all}
\end{figure*}

Putting the outlined scheme to test, we have prepended the long-term probing experiment outlined in the previous section with a dedicated provisioning phase. We captured $M=300$~ATR~responses while we randomized the server's CPU load before starting the probing attack simulation, \ie, automated needle insertion. %
The effect of the preprocessing scheme can be seen in Figure~\ref{fig:longterm_measurement_vna} where the insertion distance and long-term intra distance are shown again. In comparison to the unprocessed spectrum (see Figure~\ref{fig:longterm_measurement_vna_raw}), it becomes clear that the separability on average is increased. The distributions of insertion and intra distribution now only faintly overlap at some points in time. At these, the tampering could not be detected. However, the way of representation slightly obscures the fact the insertion distance is not i.i.d.~distributed over time and all possible holes. So far, we have aggregated all holes of the sensitive region for the insertion distance. However, given a closer look, individual holes show different levels of detectability. To see this, we have depicted the insertion distance in Figure~\ref{fig:single_drill_plots} for the three holes A, B, and C indicated in~\autoref{fig:server_areas}. These holes show clearly distinct levels of insertion distance that remain stable over time. Two of the holes can be distinguished especially well from the long-term intra distance. This reinforces the point already made in \autoref{sec:ideal_box_to_realistic}: In an actual application, one has to pay special attention to cover all sensitive parts of the system well. This does not, however, take away from the fact that tamper events can be detected well even in adverse conditions with a lot of noise, if the measurement system is well positioned and calibrated.

\subsection{Commodity UWB Measurements}
The previous results already highlight the efficacy of the ATR principle. One potential obstacle on the path to utilization, however, could be the cost and bulkiness of the measurement system. For the results described so far, we have used a professional VNA which is a great tool for experimentation due to its flexibility and low noise-floor. This instrument, however, is not a necessary requirement to implement an ATR. To show this, we have also captured responses during the long-term experiment with the UWB measurement system (connected to the antennas through an RF switch). %
The utilized UWB transceivers are both low-cost and fairly miniaturized and are thus a good candidate for an actual ATR implementation. The UWB measurement process is different from the VNA, as the wireless channel estimations are made in the time instead of frequency domain. However, they are related through the Fourier transform~\cite{goldsmith_wireless_2005} and, thus, the captured responses essentially carry similar information on the wireless propagation environment. This is visible in Figure~\ref{fig:longterm_measurement_uwb}, where it is shown that the UWB system also performs well in detecting tamper events over a long period of time. To obtain the results, we applied a selection of the channel data based on the provisioning phase as outlined in the previous section. The intra distance does not drift significantly and is mostly separated from the insertion distance. However, compared to the VNA-based measurements, see~Figure~\ref{fig:longterm_measurement_vna}, there is a lowered detection margin, \ie, a small number of holes will go undetected. 

The VNA and UWB results may also be compared quantitatively with regard to detection rates. For this, we apply a threshold to the response deviation and count the correct classifications of the respective responses. Selecting the threshold such that none of the legitimate responses is considered a tamper event, \ie, zero false positives, the UWB measurement shown in Figure~\ref{fig:longterm_measurement_uwb} yields a detection of $90$ to $108$ of all $117$ holes. In contrast, based on the results shown in Figure~\ref{fig:longterm_measurement_vna}, we were able to detect $114$ to $116$ of all $117$ holes with the VNA system. The exact numbers change over the course of time along the small measurement drift associated with the needle insertion impact.

\section{Discussion}
In this section, we discuss the experimental setup and give directions for future work.

\subsection{Experimental Setup}
We designed our experimental setup to explore the ATR capabilities and their underlying physical mechanisms. In regard to the complexity of an actual ATR deployment, we made some simplifications to facilitate experimentation.

\Paragraph{Attack Scope}
First, our setup only simulates probing attacks, \ie, we did not wiretap actual PCB traces but inserted needles into the respective test enclosures. While the setup could surely be used for actual probing attacks, we have intentionally decided against doing so because we were more interested in a broader characterization of the attacker impact instead of one specific attack. Further classes of attacks, that are based on manipulations other than the insertion of a fully metallic object, are not covered by our analysis and are an important part of future work.

\Paragraph{Premade Holes}
Our test enclosures were prepared using precision machining to drill the holes for probing into the lids. Naturally, an attacker would have to revert to existing openings in the enclosure or would even need to drill dedicated holes into a live system. The latter is already a difficult exercise on its own that we believe can have a noticeable impact on the ATR measurements. This might force an attacker to switch to more elaborate techniques such as laser drilling. Note that we did not use any additional sealing material, \ie, we assume the enclosure itself forms a sufficient boundary towards radio waves. However, an additional encapsulating wire-mesh would help to enhance the attacker's imprint on the wireless radio measurement upon insertion of a foreign object.

\Paragraph{Attack Timing}
To monitor the attacker's trace in the ATR response in a step by step manner, we have synchronized the ATR measurements with the needle positioning system, i.e., the response generation and the needle movement happen in separate consecutive steps. In an actual deployment, however, the ATR would continuously monitor the environment while an attack could occur at any time. %

\Paragraph{ATR Hardware Placement}
In our experimental ATR realization, we placed the radio measurement systems outside the protected enclosure. For an actual deployment, the measurement system needs to be installed within the enclosure in the sensitive region, so that the ATR is able to protect itself.

\Paragraph{EMI and Jamming Consideration}
Using radio propagation for tamper detection has the inherent benefit of not relying on distributed sensors and uses an open medium that is shared with other services, \eg, \mbox{Wi-Fi} or cellular communication. %
Therefore, the ATR is potentially subject to EMI from other devices emitting radio signals. During our experiments, we did not experience issues due to EMI, although the setup was located in a busy office building with considerable wireless traffic nearby. Here the metallic enclosure will provide a certain immunity against interference and the ATR system designer can additionally choose to use robust waveforms, \eg, spread-spectrum techniques, or allocate spectral regions with low signal pollution. These remarks naturally also apply to adversarial attempts of wireless jamming as a special case of interference. Note that wireless jamming against the ATR will likewise result in triggering the tamper response, so that the attacker gains no advantage.

\subsection{Future Work}
In this paper, we examined the ATR to provide a system-level tamper detection mechanism which is a central requirement of protection against physical attacks. Our findings may provide a basis for future work to further study the ATR principle. Here, we outline possible research directions. 

\Paragraph{Alternative Measurement Systems}
We demonstrated ATR implementations based on wideband channel measurements in the lower GHz frequency range gathered by a VNA and low-cost UWB transceivers. Many other measurement devices, including dedicated instrumentation as well as off-the-shelf devices are potentially well-suited to provide fine-grained radio channel measurements for ATR. Due to the high spatial resolution associated with smaller wavelengths~\cite{radarcross_Crispin1965}, future work should address the application of multi-GHz and THz waveforms that could potentially push the ATR sensitivity into regions of sub-\SI{100}{\um} resolution~\cite{sheikhOnChannelsRough}. It is worth noting that highly integrated radio systems operating above \SI{20}{\GHz} become increasingly available at low cost due to their relevance for 5G and radar applications~\cite{etsi_white_paper, gsma_whitepaper_5g}. 
One important property of a measurement system is the total amount of time it takes to capture a response. Our VNA and UWB systems, that we did not specifically optimized for speed, require \SI{250}{\ms} and \SI{700}{\ms}, respectively. The response acquisition can certainly be accelerated by additional engineering to minimize the time window available for probing attacks. %
To enhance the ATR coverage, \ie, the sensitive region, the stimulus signal could be received from multiple locations within the enclosure in parallel. Furthermore, several ATR systems could operate simultaneously by employing traditional orthogonal signaling~\cite{goldsmith_wireless_2005}.%

\Paragraph{Reference Tracking}
We compare the incurring ATR responses against an initial reference. However, this does not take into account possible aging of the environment or the ATR measurement system. Resulting slow drifts of the ATR response may be tackled by gradually renewing the reference measurement. However, more work is required to study attack vectors arising with such a mode of operation.   

\Paragraph{Anomaly Detection}
We use the mean normalized deviation between radio channel measurements to detect adversarial tamper attempts. While this simple scalar metric provides an effective solution to the task, we believe that elaborate anomaly detection approaches could further boost detection performance. For instance, we have shown that the insertion of a probing needle leaves accentuated spectral imprints (see~\autoref{fig:spectral_effect_needle}) that could potentially allow even better detection by using more refined signal processing.

\Paragraph{System Design}
Our work provides important insights to the ATR, showing that the underlying principle is capable of robust tamper detection. However, in order to deploy the ATR, system design guidelines need to be found.
This should take electromagnetic compatibility~(EMC) tests into account, ensuring that ATR signals do not interfere with the protected device and comply with regulations and shielding requirements~\cite{nsa_tempest}. 

Seeing how cumbersome experimental evaluation of tampering can be, it would be especially interesting whether an ATR can be simulated well enough that coverage of sensitive regions can be accurately predicted. There are many degrees of freedom in the engineering process, such as placement and design of antennas. In this context, simulations would help to speed up the search in the design space.

\Paragraph{Changing Environmental Conditions}
We applied different load states to our test server that in turn lead to temperature swings by more than $20^{\circ}$C of the CPU temperature. This, however, does not cover the full conceivable range of environmental conditions a system might experience. For example, if a system is initialized after manufacturing and then enters the supply chain it will probably be subject to much lower and higher temperatures, changes in humidity, vibrations, and mechanical shocks. 
For an actual deployment, these factors have to be taken into account to determine a fitting threshold for tamper detection.

\Paragraph{Adaptive Attacks}
In our experiments, we examined different challenges that one encounters in designing an ATR system. The goal was not to propose a concrete fully-fledged system that is ready to be applied in the real world. Consequently, the attacks that we show are supposed to highlight what can be detected by the system. Building upon our work, the next logical step would be to design a concrete system with a clear definition of what constitutes a break of its security. This system could then be put to the test, possibly by a third party that tries its best to find a working attack. Doing this, they could take full knowledge of the defensive system into account. This includes, for example, the apparent weakness in detecting tampering in close vicinity to the enclosure boundary, as discussed in \autoref{sec:ideal_box_to_realistic}.

Another idea worth investigating is to replace the metallic probing needle with an RF-transparent material such as low-loss plastic. 
This could facilitate probing attempts, with the needle now acting as a rigid support for a very thin wire\footnote{Wires of a diameter of less than$~\SI{0.05}{\mm}$ are commercially available.}. %
Alternatively, one could mount a small conductive tip on an otherwise RF-transparent needle, which would be useful in attacks that require the shortening of electrical contacts. This general attack scenario limits the amount of material that is visible for the ATR and therefore warrants a careful analysis.

\section{Related Work}
Design and evaluation of anti-tamper solutions are usually done by companies and less often by academia. Consequently, public sources on this topic are rather sparse. 
One starting point is the already mentioned analysis of the IBM~4758 by Anderson~\cite{andersonSecEngineering} which gives insight on a security-mesh based design. The assembly process of its successor, the IBM~4765, is outlined by Isaacs~\etal~\cite{isaacs2013tamper}. In contrast,  Obermaier and Immler~\cite{obermaierPresentFuturePhysical2018} disassembled two HSMs to present an analysis of the implemented anti-tamper measures and reason about potential weaknesses. %

While successful real-world attacks on state-of-the-art HSMs are, to the best of our knowledge, not publicly known, several works present attacks on devices implementing anti-tamper measures. Early work by Anderson and Kuhn~\cite{andersonTamperCaution1996} outlined invasive attacks against supposedly tamper resistant chips. Helfmeier~\etal~\cite{helfmeierBreakingEnteringSilicon2013} circumvent the tamper detection mechanism of integrated circuits through microprobing attacks from the chip backside. Drimer~\etal~\cite{drimerThinkingBoxSystemLevel2008} demonstrated needle probing attacks to circumvent the anti-tamper mechanisms of two PIN entry devices for electronic payment systems, enabling extraction of card information and PIN codes. Weingart~\cite{weingartPhysicalSecurityDevices2000a} surveyed physical attacks and countermeasures, notably including a brief paragraph on the use of microwaves, although without giving further details or references. 

Apart from examination of attacks on commercially available designs, different novel approaches for anti-tamper protection have been proposed. Xu~\etal recently suggested impedance variations of bus lines between a CPU and peripheral chips to be used for detection of tampering such as probings~\cite{xuBusAuthenticationAntiProbing2020a}. A creative solution was put forward by G{\"o}tte and Scheuermann~\cite{cryptoeprint:2021:055} that consists of constantly rotating HSMs at more than \SI{500}{rpm} to hamper accessibility.

Directly regarding our work, we have already mentioned in~\autoref{sec:system_idea} that using characteristics of radio wave propagation to sense environmental states in a security context has been pursued before. DeJean and Kirovski~\cite{dejeanRFDNARadioFrequencyCertificates2007} used unique radio wave characteristics in near-field measurements to authenticate hardware tags for anti-counterfeiting applications. Bagci~\etal~\cite{bagciUsingChannelState2015} used channel state information to detect tampering of IoT devices, considering device movement or rotation as tamper events. To distinguish such from environmental changes, multiple receivers must simultaneously receive signals from the protected device. In a talk, Zenger~\etal~\cite{enc_puf_36c3} proposed to leverage radio wave propagation within a small enclosure as a PUF, detecting insertions of foreign objects. In their live demo, the sensitivity of the measurement against integrity violations was demonstrated. However, experimental results especially with regard to reproducible tampering attempts were not given.

Physical unclonable functions (PUFs) are an interesting security primitive~\cite{maesPufOverview}, that can, for example, be used to establish authenticity or to derive cryptographic keys from their physical structure. It is generally assumed that invasive attacks on PUFs are futile because these destroy the physical structure to a degree that the true PUF output cannot be recovered anymore.
Based on this idea, PUF designs that are explicitly geared towards anti-tamper applications have been proposed. One example is given by Tuyls~\etal~\cite{tuylsReadProofHardwareProtective2006} who proposed a protective layer with PUF properties to cover integrated circuits.
Another example comes from Immler~\etal~\cite{immlerSecurePhysicalEnclosures2018} who extended the idea of security meshes and presented a protective cover with PUF properties to protect PCB modules against physical access. The PUF response is based on measurements of capacitances that are present within the mesh. It is used to derive a cryptographic key that is bound to the the physical integrity of the system. These covers are a promising primitive but cannot be easily extended to system-level protection which is a drawback compared to the ATR principle. In principle, the ATR could also be framed as a PUF. If, however, the goal is to derive a cryptographic key, one has to determine and analyse a source of entropy, which is non-trivial.

\section{Conclusion}

In this paper, we experimentally examined the capabilities of a prototypical radio-wave based tamper detection system that we term anti-tamper radio~(ATR). Bringing the inherent advantages of wireless radio systems to the field of anti-tamper solutions, the ATR enables the detection of tamper events for assets with complex geometry and larger dimensions. To this end, we leverage the propagation characteristics of radio waves within the protected environment as a distributed sensing modality, that is able to indicate physical attacks in the form of integrity violations by foreign objects. Using a robotic needle positioning system, we simulated probing attacks against two ATR-protected environments. Beginning with an idealized enclosure, we have shown the sensitivity of our ATR implementation against penetration with needles of \mbox{sub-mm} diameter. We then successfully demonstrated the detection of probing attacks on a running 19\inch~server in a long-term experiment for $10$~days. Furthermore, we have shown the feasibility to implement ATR even with cheap off-the-shelf measurement systems.

Due to its unique properties, the ATR could contribute to the proliferation of protection against physical attacks, even for low-cost devices. Thus, we hope that our work will stimulate further interest in radio wave-based anti-tamper solutions.

\section*{Acknowledgements}
We thank Simon Mulzer for his help with programming the Cartesian robot. This work was supported by the Deutsche Forschungsgemeinschaft (DFG, German Research Foundation) under Germany’s Excellence Strategy - EXC 2092 CaSa - 390781972.

\bibliographystyle{IEEEtranS}
\bibliography{probing_refs}

\appendix

\section{Appendix}

\subsection{Lid opening}
\begin{figure}[h]
\centering
\includegraphics[width=0.99\columnwidth]{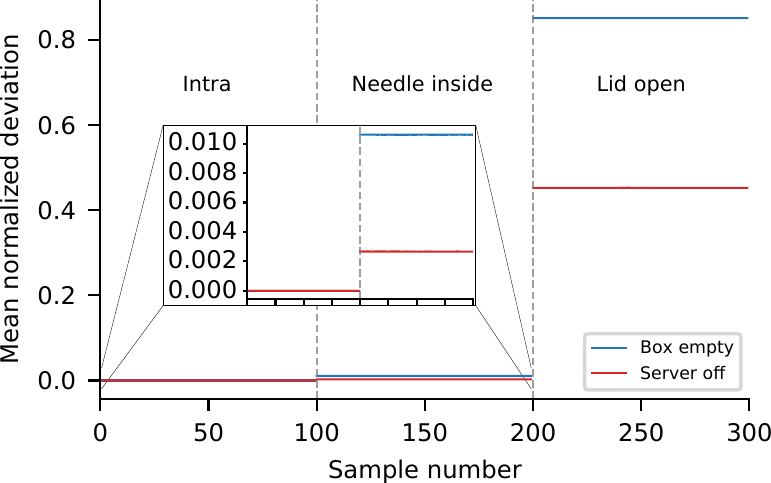}
\caption{MND of ATR responses (referred to an initial reference measurement) for the empty aluminum box and the server enclosure. Samples $0$ to $99$: without tampering (Intra). Samples $100$ to $199$: tampering with a \SI{1}{\mm} diameter needle inserted \SI{40}{\mm} deep into one drill. Samples $200$ to $299$: top lid removed. The effect of the lid removal is by far greater than that of the needle insertion.} %
\label{fig:lidopening}
\end{figure}

In our experiments that are described in Section~\ref{sec:probing_ideal_box}, we have shown that the sensitivity of detecting needles scales with the needle size and insertion depth. This directly suggests that larger physical changes result in stronger response deviation. To validate this, we performed an experiment in which the effect of needle insertion is compared with the removal of the enclosure lid. As shown in Figure~\ref{fig:lidopening}, it is indeed the case that the lid removal has a by far greater effect on the response than the needle insertion.
\FloatBarrier

\subsection{Needle electrical potential}
\begin{figure}[h!]
\centering
\includegraphics[width=0.99\columnwidth]{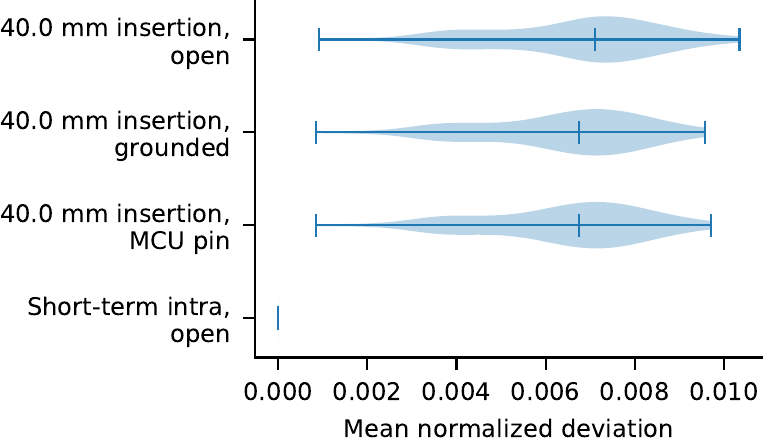}
\caption{Short-term intra distance and impact of \SI{40}{\mm} insertion of an \SI{1}{\mm} diameter needle being connected to a microcontroller~(MCU) pin with activated pull-up resistor, grounded, and open-circuit. The needle was inserted into equidistant holes across the empty aluminium box.}
\label{fig:needle_potential}
\end{figure}

Additionally to varying the size of physical perturbation from the probing needle, we also tested how varying electrical potentials affect the ATR response. In our experiment, we inserted a probing needle of \SI{1}{\mm} diameter \SI{40}{\mm} deep into $110$~holes of the empty aluminium box. We repeated this procedure with the needle being open-circuit, grounded, and connected to a microcontroller pin. Figure~\ref{fig:needle_potential} depicts the distribution of insertion distances and also shows the intra distance. As we can see, the needle detectability does not significantly change.

\end{document}